**А.В. ПАЛАГИН**, академик НАНУ, д-р техн. наук, профессор, заместитель директора,
palagin_a@ukr.net

**Н.Г. ПЕТРЕНКО**, д-р техн. наук, ведущий науч. сотр.,
petrng@ukr.net

**К.С. МАЛАХОВ**, младш. науч. сотр.,
malakhovks@nas.gov.ua

Институт кибернетики имени В.М. Глушкова НАН Украины, просп. Глушкова, 40, Київ 03187, Україна.

# ИНФОРМАЦИОННАЯ ТЕХНОЛОГИЯ И ИНСТРУМЕНТАЛЬНЫЕ СРЕДСТВА ПОДДЕРЖКИ ПРОЦЕССОВ ИССЛЕДОВАТЕЛЬСКОГО ПРОЕКТИРОВАНИЯ *SMART*-СИСТЕМ

*Рассмотрены методологические основы построения информационной технологии, формализованных моделей и инструментальных средств реализации процессов исследовательского проектирования Smart-систем, основанные на использовании концепций трансдисциплинарности и онтологического управления. Предназначены для построения систем работы со знаниями, научно-технического творчества и разработки объектов новой техники.*

Уровень развития интеллектуальных информационных технологий в значительной мере влияет на эффективность процессов, происходящих в экономической, научно-технической, образовательной и других сферах деятельности общества. Процессы глобальной информатизации мирового сообщества ориентированы, прежде всего, на построение общества знаний и носят ярко выраженный трансдисциплинарный характер. Несомненный лидер при этом — технологии инженерии знаний, в том числе ее новое направление — онтологический инжиниринг. Посредством этих технологий реализуют процессы управления знаниями (*Knowledge Management*), и успехи в этом направлении во многом определяются уровнем интеллектуализации и эффективности компьютерных систем [1–5]. В настоящее время наметилась активизация научных исследований на стыке разных предметных дисциплин (междисциплинарные исследования), и в так называемых кластерах конвергенции (трансдисциплинарные исследования). Поддержке этих исследований способствуют построение знание-ориентированных информационных систем, совершенствование процессов исследовательского проектирования (ИП), разработка методов и инструментальных средств онтологического анализа естественно-языковых объектов с целью извлечения из них знаний, прикладных аспектов применения онтологий, в частности при разработке электронных учебных курсов (и других аспектов электронного образования), метаонтологий, систем интеграции знаний в трансдисциплинарных кластерах конвергенции и пр.

ИП как разновидность научных исследований характеризуется тем, что основные его этапы связаны с процессом описания предполагаемого образа проектируемого объекта новой техники (ОНТ) [6], который строится как ряд интерактивных процедур более глубокого погружения в описание объекта новой техники и формирования промежуточных вариантов технического решения. При этом преобразование описания объекта проектирования зависит как от знаний о предметной области (ПдО) в целом, так и от знаний и опыта проектировщика.

Процесс ИП в обобщенном виде [7] состоит из следующих этапов:

- сбора материала, представляющего ПдО;
- формирования цели исследовательского проекта;
- анализа материала и онтологического описания ПдО;
- выявления противоречий и формулирования проблемной ситуации;
- постановки задачи исследовательского проекта;
- уточнения проблемной ситуации;
- выявления аналогов-прототипов, формирования совокупности их технических признаков;
- формирования совокупности технических признаков и облика ОНТ;
- выполнения эскизно-технического этапа проектирования и подготовки материалов к патентованию.

*Цель статьи* — разработка методологических основ построения систем исследовательского проектирования объектов новой техники (как разновидности *Smart*-систем) с использованием онтологической концепции и технологии *Smart Research & Development*, содержащих модели, информационную технологию и инструментальные средства когнитивных процессов обработки информации.

## *Smart*-системы

Процесс интеллектуализации в области средств информатики и вычислительной техники вышел на новый масштабный уровень, расширив пространство и функциональность их приложений в современном обществе, чему в значительной степени способствовала стратегическая рамочная программа «Горизонт 2020». Она открыла эру интенсивного развития *Smart*-систем различного уровня и назначения [8]. Это направление поддержано совместной технологической инициативой ряда европейских международных организаций. Многолетний стратегический план реализации программы исследований и инноваций в области электронных компонент, систем и технологий (*MASRIA*) обеспечивает развитие следующих основных функциональных доменов: *Smart-society*, *Smart-mobility*, *Smart-energy*, *Smart-health*, *Smart-production*. По сути, речь идет о технологиях, воздействующих на все аспекты функционирования современного общества, таких как взаимодействие в реальном времени между машинами, людьми и объектами окружающего мира; обеспечение безопасности личности и общества; эффективное снабжение и распределение (вода, продукты питания и др.); логистика; смарт-администрирование и в целом поддержание устойчивого развития общества. Все множество перечисленных задач и областей деятельности — объекты исследования и разработок стратегического плана *MASRIA*. Отметим, что в «Концепції розвитку електронного урядування в Україні» и цифровой экономики предусмотрено применение современных инновационных подходов, методологий и технологий, в частности Интернет-вещей, облачной инфраструктуры, *Blockchain*, *Big Data* и др. [9]. Кассификация атрибутов, характеризующих *Smart*-системы и их краткое описание [10], представлена в таблице.

Понятие *Smart*-система в онтологической иерархии тесно связано на верхнем уровне с такими понятиями, как кибер-физические системы и другими разделами *Smart*-общества. На нижних уровнях расположены различные *Smart*-устройства, обеспечивающие реализацию приложений пользователей. Анализ публикаций в зарубежных и отечественных научных изданиях позволил синтезировать несколько верхних уровней онтологии ПдО *Smart*-система [8, 11].

| Атрибут | Описание |
|---|---|
| Адаптация, приспособление | Способность изменять физические или поведенческие характеристики в соответствии с изменениями в окружающей среде или выживания в ней |
| Восприятие, считывание, распознавание, понимание | Способность идентифицировать, распознать, понять и/или осознать феномен, событие, объект, воздействие |
| Логический вывод | Способность делать логический вывод (ы) на основе исходных данных, обработанной информации, наблюдений, доказательств, предположений, правил и логических рассуждений |
| Обучение, освоение | Способность приобретения новых или изменения существующих знаний, опыта, поведения для повышения производительности, эффективности, навыков |
| Антиципация (предугадывание событий) | Способность мыслить или рассуждать, чтобы предсказывать события или собственные действия |
| Самоорганизация и реструктуризация (оптимизация) | Способность системы изменять свою внутреннюю структуру (компоненты), самовосстанавливаться и самоподдерживаться целенаправленным (неслучайным) способом при соответствующих условиях, но без внешнего агента/сущности |

Онтографическое представление понятия *Smart*-система приведено на рис. 1.

Известны следующие определения кибер-физических систем и *Smart*-устройств [8].

***Кибер-физические системы*** – это электронные системы, компоненты и программное обеспечение, тесно взаимодействующие с физическими системами и их окружающей средой: встроенный интеллект предоставляет возможности для отслеживания, мониторинга, анализа и управления физическими устройствами, компонентами и процессами в различных областях применения. Их способность подключаться и взаимодействовать через все виды сетей и протоколов (включая Интернет, проводную и беспроводную связь) позволяет им координировать и оптимизировать функциональные возможности физических систем.

Сегодня общепринятого определения *Smart-системы* нет. По мнению авторов, корректным будет следующее определение.

*Smart*-системы — это системы, которые могут функционировать как автономно, так и в составе более сложных, кибер-физических систем, и способны обеспечить:

- взаимную адресацию и идентификацию;

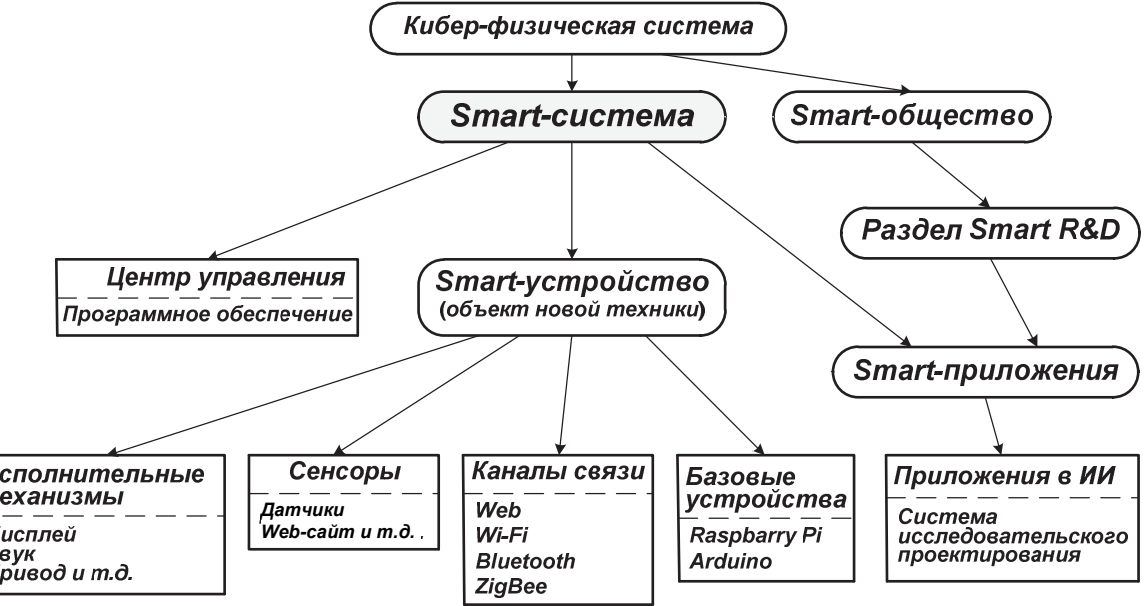

***Рис. 1.*** Онтографическое представление понятия *Smart*-система

- диагностирование, описание и квалификацию окружающей среды;
- прогнозирование событий, принятие решений и их исполнение;
- наличие более одной коммуникационной технологии.

*Smart*-устройство — это физический объект, в цифровом формате взаимодействующий с одним или несколькими объектами:

- сенсоры (температура, свет, движение);
- исполнительные механизмы (дисплеи, звук, двигатели);
- управляемое вычисление (может запускать программы и логику);
- интерфейсы связи.

Сформулированные в [7] основные принципы комплексного подхода к решению проблемы разработки *Smart*-систем для потребностей современного общества в конкретной реализации и с учетом технологий *Internet* и *Semantic Web* трансформируются в следующие:

- обработка знаний — при проектировании ОНТ необходимо иметь знания нескольких предметных областей, формально описанных на одном из языков, характерных для *Semantic Web* (например, *RDF*);
- трансдисциплинарность — системная интеграция знаний задействованных предметных областей;
- онтологическая концепция — структура знаний ПдО представлена в виде онтологий с разделением на статическую и динамическую составляющие;
- определение проблемной ситуации и формулирование проблемы;

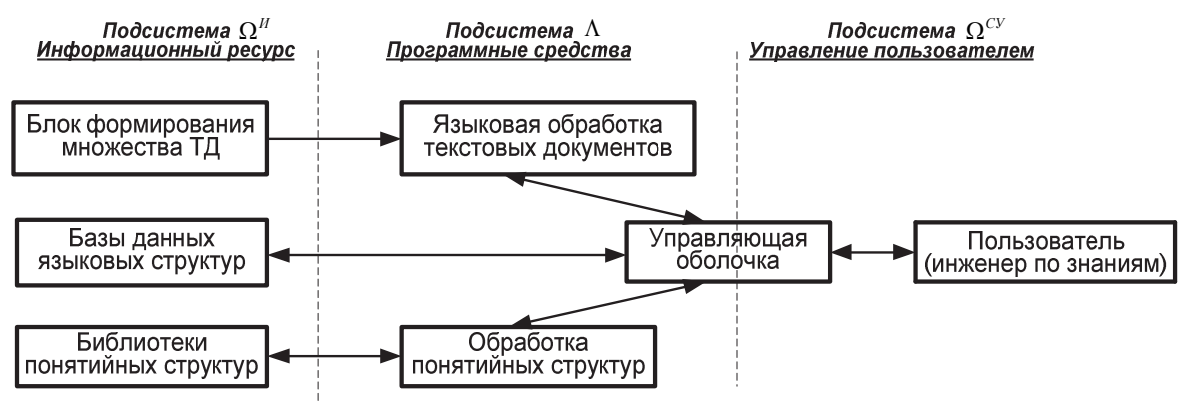

***Рис. 2.*** Обобщенная блок-схема ИКОН

• виртуализация, унификация и стандартизация технических решений — коммуникация *Smart*-систем между собой и с пользователями осуществляется через *Web*-среду и/или «Облако».

## Инструментальные средства поддержки процессов ИП

Сложность выполнения современных научных исследований, в том числе и *Smart R&D*, предъявляет более высокие требования к инструментальным средствам их поддержки:

• оперировать большими объемами текстовой информации с целью извлечения из них предметных знаний;

• формировать концептуальные структуры знаний (онтологии) как на стыке предметных дисциплин, так и в кластерах конвергенции;

• системно интегрировать построенные онтологии с целью обнаружения новых знаний. Такой проблемной ориентации соответствует инструментальный комплекс онтологического назначения (ИКОН) [5].

ИКОН реализует ряд компонентов единой информационной технологии (ИТ):

• поиск в сети *Internet* и/или в других электронных коллекциях текстовых документов (ТД), релевантных заданной ПдО, их индексация и сохранение в базе данных;

• автоматическая обработка естественно-языковых текстов;

• извлечение из множества ТД знаний, релевантных заданной ПдО, их системно-онтологическая структуризация и формально-логическое представление на одном или нескольких общепринятых языках описания онтологий (*Knowledge-Representation*);

• создание, накопление и использование больших структур онтологических знаний в соответствующих библиотеках;

• системная интеграция онтологических знаний как одна из основных компонент методологии междисциплинарных и трансдисциплинарных научных исследований.

ИКОН состоит из трех подсистем и представляет собой интеграцию разного рода информационных ресурсов, программных средств обработки и процедур пользователя, которые, взаимодействуя, реализуют совокупность алгоритмов автоматизированного итерационного построения понятийных структур предметных знаний, их накопления и/или системной интеграции [5, 12].

Обобщенная блок-схема ИКОН представлена на рис. 2.

## Модели процессов технологии автоматизированного построения онтологий ПдО

Методика проектирования ИКОН предполагает разработку, в том числе информационной, функциональной и поведенческой моделей, описывающих различные аспекты его функционирования, обработки и хранения информации.

Под *функциональной моделью* ИКОН предполагается описание совокупности выполняемых инструментальным комплексом функций, характеризующее состав функциональных блоков и их взаимосвязи.

Под *информационной моделью* ИКОН понимается отражение отношений между блоками в виде структур данных и информационных потоков, их состав и взаимосвязи.

Под *поведенческой моделью* ИКОН понимается описание информационных процессов или динамики функционирования. В ней описаны состояния инструментального комплекса, события, переходы (и условия переходов) из одного состояния в другое и последовательность событий.

Модель информационной технологии представляется четверкой [12]:

$$M =< P, A, X, \Omega >,$$

где $P = \{ p_i \}, i = \overline{1,n}$ — множество процессов, реализующих ИТ; $A = \{ A_j \}, j = \overline{1,m}, m \geq n$ — множество алгоритмов, реализующих множество процессов $\{p_i\}$, причем может быть несколько алгоритмов, реализующих некоторый процесс $p_i$; $X$ — множество сущностей, описывающих заданную ПдО и участвующих в реализации алгоритмов $\{ A_j \}$; $\Omega$ — обобщенная архитектура инструментальных средств, участвующих в реализации ИТ, которая, в свою очередь, описывается тройкой:

$$\Omega =< \Omega^{И}, Л, \Omega^{СУ} > \qquad (1)$$

и проектируется в соответствии с *онтологическим* методом и моделью проектирования архитектуры знание-ориентированной информационной системы с онтолого-управляемой архитектурой [5], где: $\Omega^{И}$ — подсистема «Информационный ресурс»; $Л$ — подсистема программных средств, реализующих ИТ; $\Omega^{СУ}$ — подсистема управления множеством процессов $\{p_i\}$. Она включает также процедуры пользователя, такие как настройка инструментария на заданную ПдО, определение необходимого и достаточного объема базы знаний (уровней иерархии онтологии ПдО), принятие решений о завершении итеративных циклов в общем алгоритме построения онтологии и др.

Рассмотрим подсистемы (1) инструментального комплекса.

Подсистема $\Omega^{И}$ «***Информационный ресурс***» состоит из блоков формирования лингвистического корпуса текстов, базу данных языковых структур и библиотек понятийных структур. Первый компонент представляет собой различные источники текстовой информации, поступающей на обработку в систему. Второй компонент представляет собой различные базы данных обработки языковых структур, часть из которых формируется (наполняется данными) в процессе обработки ТД, а другая — формируется до процесса построения онтологии ПдО и, по сути, является электронной коллекцией различных толковых словарей. Третий компонент представляет собой совокупность библиотек понятийных

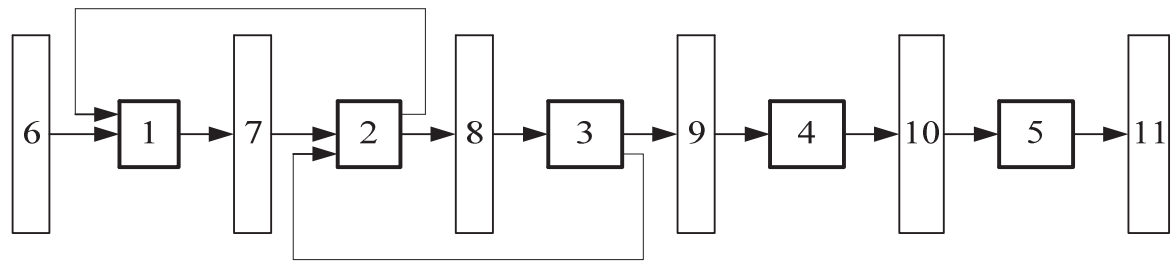

***Рис. 3.*** Блок-схема интегрированной информационной технологии структур разного уровня представления (от наборов терминов и понятий до высокоинтегрированной онтологической структуры междисциплинарных знаний) и результат реализации некоторого проекта (проектирования онтологии ПдО).

Подсистема $Л$ «***Программные средства***» включает в себя блоки обработки языковых и понятийных структур и управляющую графическую оболочку. Последняя во взаимодействии с инженером по знаниям осуществляет общее управление процессом реализации связанных информационных технологий.

Подсистема управления $\Omega^{СУ}$ осуществляет подготовку и реализацию процедур предварительного этапа проектирования, а на протяжении всего процесса осуществляет контроль и проверку результатов выполнения этапов проектирования, принимает решение о степени их завершенности и в случае необходимости — о повторении некоторых из них.

Для разработки моделей технологии автоматизированного построения онтологий ПдО необходимо:

- разработать информационную модель множества процессов $P$, составляющих ИТ;
- разработать функциональную модель ИТ;
- разработать подсистему программно-аппаратных средств $Л$, реализующих множество алгоритмов $A$;
- при решении общей задачи построения инструментария учитывать критерии эффективности, в частности, уровень автоматизации построения онтологической базы знаний ПдО и ограничение реального времени получения результата.

***Информационная модель ИТ***

На рис. 3 представлена блок-схема интегрированной информационной технологии (или информационной модели) автоматизированного построения онтологий предметных областей, где приняты следующие обозначения:

1 — технология поиска в различных источниках ТД, релевантных к заданной ПдО;

2 — технология автоматического лингвистического анализа ТД, описывающих заданную ПдО. В случае разрешения грамматической омонимии (при необходимости) осуществляется возврат к блоку 1 для поиска соответствующей текстовой информации в лингвистических словарях;

3 — технология извлечения из множества ТД онтолого-информационных структур и их хранение в соответствующих базах данных. В случае разрешения лексической неоднозначности (при необходимости) осуществляется возврат к блоку 2 для построения дополнительных синтактико-семантических структур;

4 — технология формально-логического представления и интеграции онтологических структур в онтологическую базу знаний ПдО;

5 — технология обработки и управления данными и знаниями большого объема;

6 — источники ТД (сеть Интернет, монографии, научно-технические статьи, электронные коллекции ТД и др.);

7 — лингвистический корпус текстов, описывающий заданную ПдО;

8 — синтаксические и поверхностно-семантические структуры ТД;

9 — множество онтолого-информационных структур как результат обработки ТД;

10 — онтологическая база знаний заданной ПдО, множества терминов и понятий ПдО;

11 — библиотеки онтологий ПдО, базы данных терминов, понятий и ТД предметных областей.

Далее, в процессе разработки ИТ выполняется декомпозиция общей информационной модели на ее составляющие информационные модели, соответствующие отдельным ИТ, т.е. формируется иерархическая структура.

Описание приведенных выше блоков представлено в [5].

***Функциональная модель процессов, реализующих ИТ***

*UML*-технологии стали основой для разработки и реализации многих инструментальных средств: средств визуального и имитационного моделирования, а также *CASE*-средств различного целевого назначения. Более того, заложенные в языке *UML* потенциальные возможности могут быть использованы не только для объектно-ориентированного моделирования систем, но и для представления знаний в интеллектуальных системах, которыми, по сути, есть перспективные сложные программно-технологические комплексы [13]. Язык *UML* имеет ряд преимуществ перед другими языками и методологиями моделирования сложных программных систем, поэтому он принят за основу при разработке формальных моделей процессов, реализующих автоматизированное построение онтологий предметных областей.

Функциональная модель (включая и элементы поведенческой модели) ИТ представляет собой набор *диаграмм трех видов* [14]:

- диаграмма вариантов использования;
- диаграмма активности;
- диаграмма классов.

Цель разработки *диаграммы вариантов использования* есть:

- определить общие границы и контекст моделируемой ПдО на начальных этапах проектирования системы;
- сформулировать общие требования к функциональному поведению проектируемой системы;
- разработать исходную концептуальную модель системы для ее последующей детализации в форме логических и физических моделей;
- подготовить исходную документацию для взаимодействия разработчиков системы с ее заказчиками и пользователями.

Интерпретация *диаграммы вариантов использования*, следующая: проектируемая система представляется в виде множества сущностей или акторов (*actor*), взаимодействующих с системой с помощью так называемых вариантов использования. При этом актором или действующим лицом называется любая сущ-ность, взаимодействующая с системой извне. Это может быть пользователь, компьютерная программа или любая другая система, служащая источником воздействия на моделируемую систему. В свою очередь, вариант использования служит для описания сервисов, которые система предоставляет актору.

С помощью *диаграммы активности* изучается поведение системы с использованием моделей потока данных и потока управления. Диаграмма активности отличается от блок-схемы, описывающей только шаги алгоритма, и имеет более широкую нотацию. Например, в ней можно указывать состояния объектов.

*Диаграмма классов* описывает структуру объектов ИТ: их индивидность, отношения с другими объектами, атрибуты, функции и процедуры. Модель классов создает контекст для диаграмм состояний и взаимодействия.

Перечисленный набор диаграмм трех видов представлен в работах [12, 14, 15].

## Исследовательское проектирование систем научно-технического творчества

В качестве конкретной реализации рассмотрим *Smart*-систему, ориентированную на научно-техническое творчество на примере проектирования лингвистического процессора.

Большинство этапов (исключая интуитивные процессы творчества) могут поддерживаться информационными технологиями, основанными на онтологическом подходе и концепции семантического пространства. В соответствии с приведенными принципами необходимо построить онтологии предметных областей: «Научно-техническое творчество», в частности «Изобретательство и патентование», «Вычислительная техника», «Компьютерная лингвистика». Онтология первой ПдО рассматривается далее, а онтологии второй и третьей ПдО представлены в [16].

На рис. 4 представлена функциональная схема онтологии «Научно-техническое творчество», где приняты следующие обозначения:

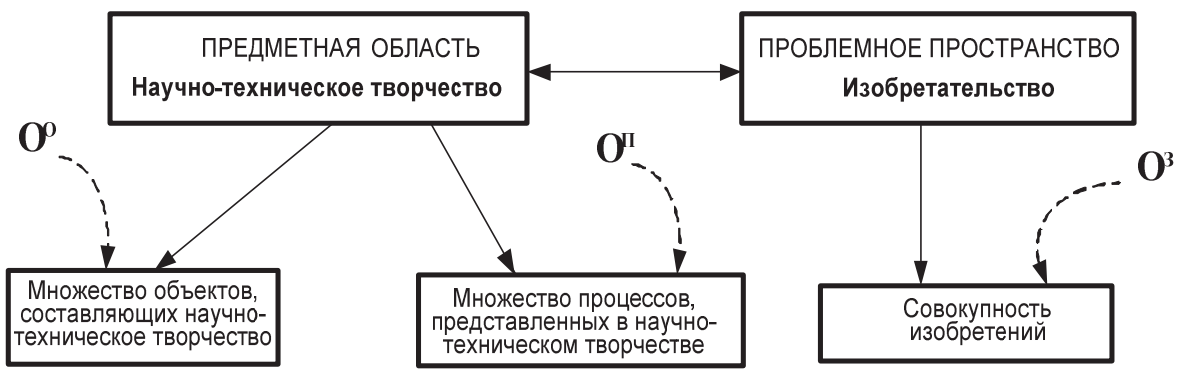

***Рис. 4.*** Функциональная схема ПдО «Научно-техническое творчество»

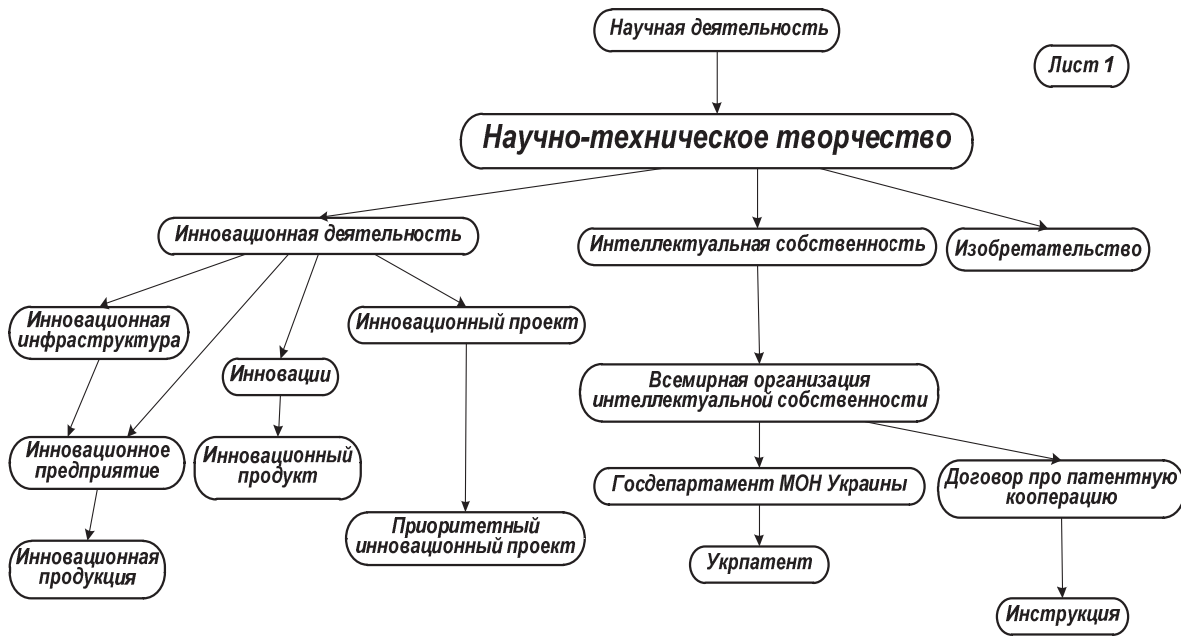

***Рис. 5.*** Верхний уровень онтологии ПдО «Научно-техническое творчество»

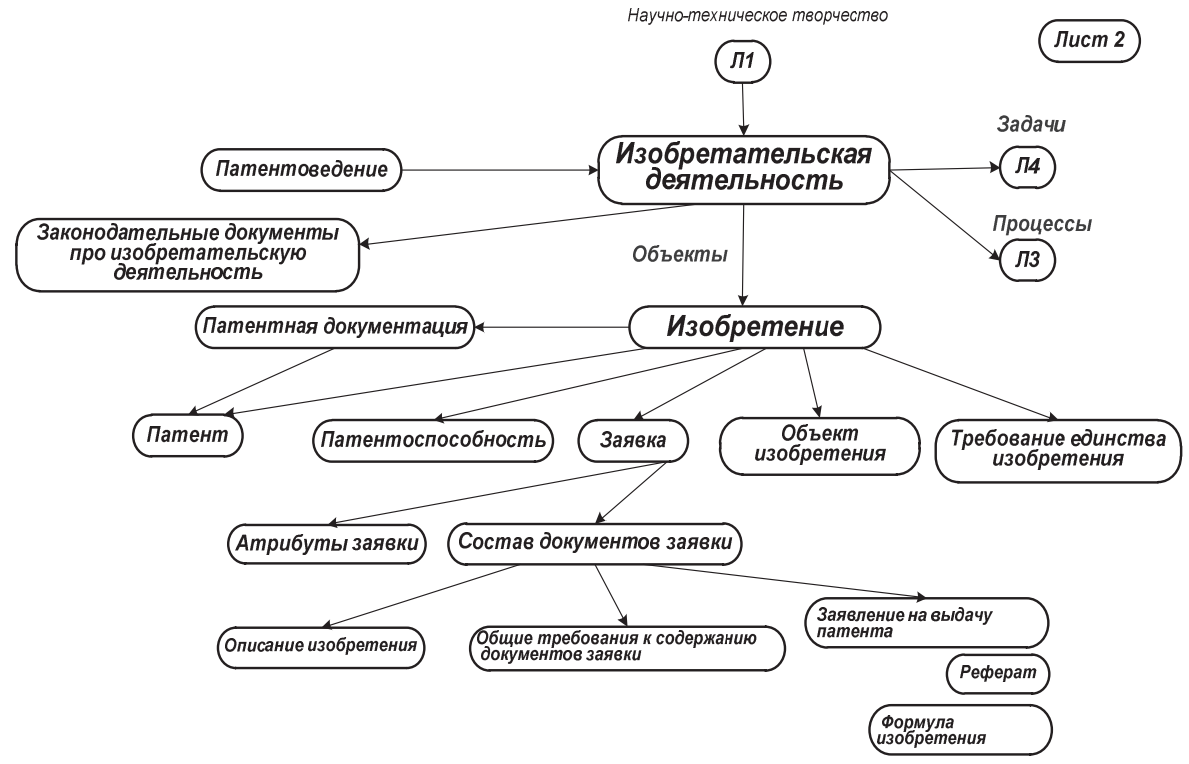

***Рис. 6.*** Онтологии ПдО «Изобретательская деятельность» и «Патентование»

- $О^О$ – онтология множества объектов (понятий, концептов) ПдО «Научно-техническое творчество», в частности «Изобретательство»;
- $О^П$ – онтология множества процессов ПдО «Научно-техническое творчество», в частности «Изобретательство», которая рассматривается как иерархическая структура процессов, подпроцессов, действий и операций;
- $О^З$ – онтология совокупности задач (изобретений), которые могут быть поставлены и решены в конкретной ПдО, рассматривается как иерархическая структура задач, подзадач, процедур и операторов.

Безусловно, приступая к созданию своего первого изобретения, научный работник или исследователь должен построить онтологии ПдО, входящие в сферу его интересов, и выполнить их системную интеграцию. Инструментом поддержки указанных процедур может быть автоматизированное рабочее место (АРМ) научного работника [17, 18].

Знания блоков $О^{О}$ и $О^{П}$ (рис. 4) изменяются не часто (в сравнении с знаниями $О^{З}$), поэтому они названы статическими и им свойственно многоразовое использование.

Знания блока $О^{З}$ представляют собой, по сути, алгоритм реализации конкретного (или типового набора) изобретения. Поэтому они названы динамическими.

На рис. 5 представлен верхний уровень онтологии «Научно-техническое творчество».

На рис. 6 в обобщенном виде представлены системно интегрированные онтологии ПдО «Изобретательская деятельность» и «Патентование». Сюда же входят объекты изобретательской деятельности (на рис. 4 обозначены как $О^{О}$), участвующие в процессе создания и оформления изобретения.

На рис. 7 в обобщенном виде представлена онтология процессов ПдО «Изобретательская деятельность» и «Патентование» (на рис. 4 обозначена как $О^{П}$), участвующих в создании и оформлении изобретения на устройство. В качестве устройства рассматривается аппаратный лингвистический процессор обработки естественно-языковых текстов.

На рис. 8 представлена онтология задач (обозначена как $О^{З}$ на рис. 4) создания и оформления изобретения на устройство «Лингвистический процессор обработки естественно-языковых текстов» [19, 20].

Кратко опишем решение задачи создания и патентования изобретения на устройство. Оно опирается на взаимодействие онтологий ($О^{О}$ $О^{П}$ и $О^{З}$), представленных соответственно на рис. 6—8. В $О^{З}$ (рис. 8) описан алгоритм, блоки которого представляют собой ряд последо-

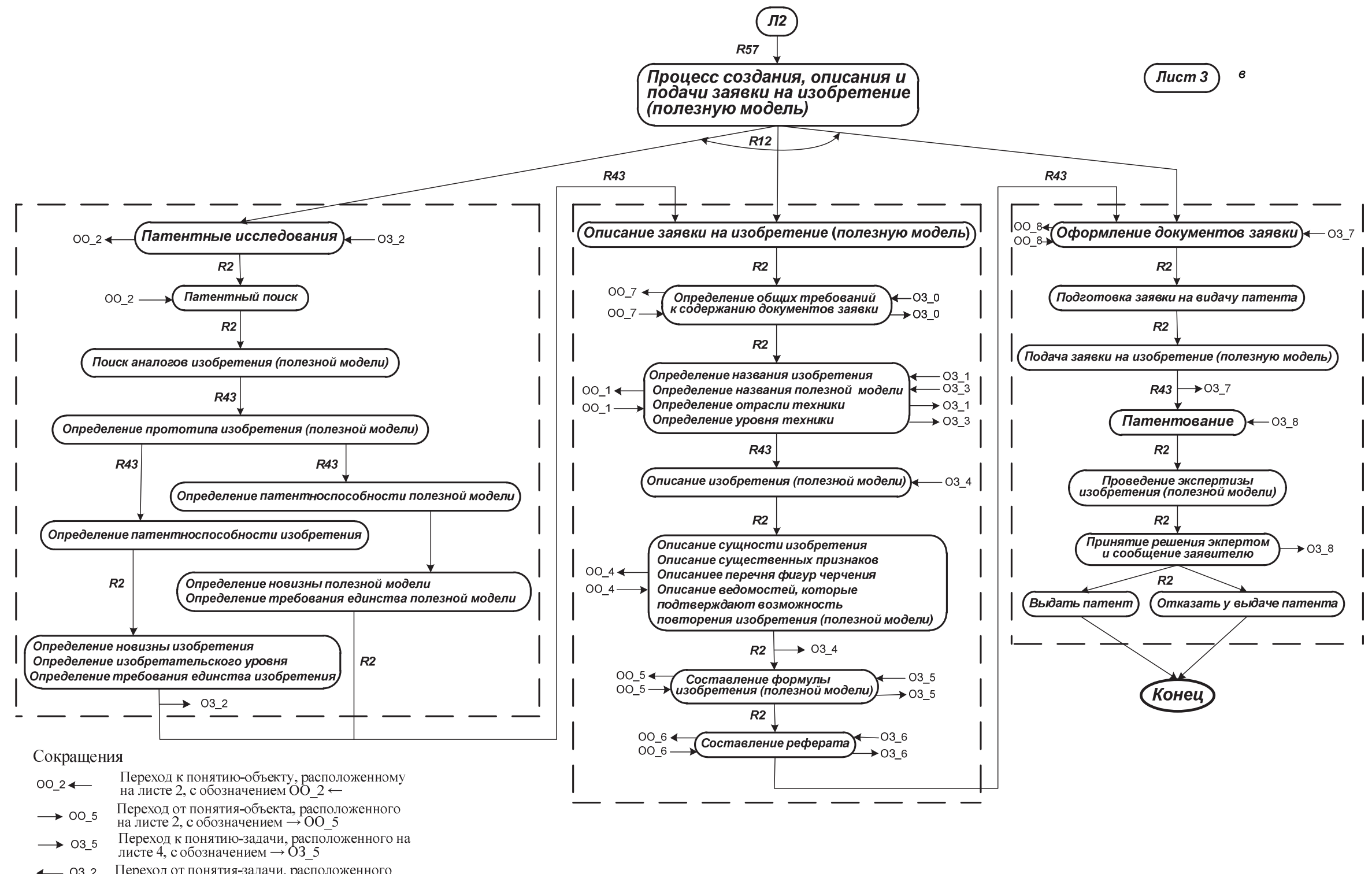

***Рис. 7.*** Онтология процессов ПдО «Изобретательская деятельность» и «Патентование»

вательных подзадач, кроме первого, в котором формируются исходные данные (тип устройства, предварительные название изобретения и средства его реализации и т.п.).

Рисунки 5–8 обозначены как Лист 1, Лист 2, Лист 3 и Лист 4 соответственно. Между блоками подзадач онтологии задач (Лист 4, рис. 8), блоками подпроцессов онтологии процессов (Лист 3, рис. 7) и понятиями онтологии объектов (Лист 2, рис. 6) существуют взаимосвязанные переходы (указаны на рисунках). Сущность таких переходов следующая.

Некоторый блок, обозначающий подзадачу в общей задаче патентования устройства (Лист 4), обращается к соответствующему подпроцессу (Лист 3), который, в свою очередь, обращается (использует для обработки) к соответствующему понятию онтологии объектов (Лист 2). Далее переходы инициируются в обратном порядке таким образом, что в итоге блок подзадачи получит результат ее решения. И алгоритм инициирует решение следующей подзадачи.

Описанная выше онтологическая составляющая — ядро системы ИП в области научно-технического творчества, и в частности – изобретательской деятельности. Такая система повышает эффективность и ускоряет процесс изобретательской деятельности [19, 20].

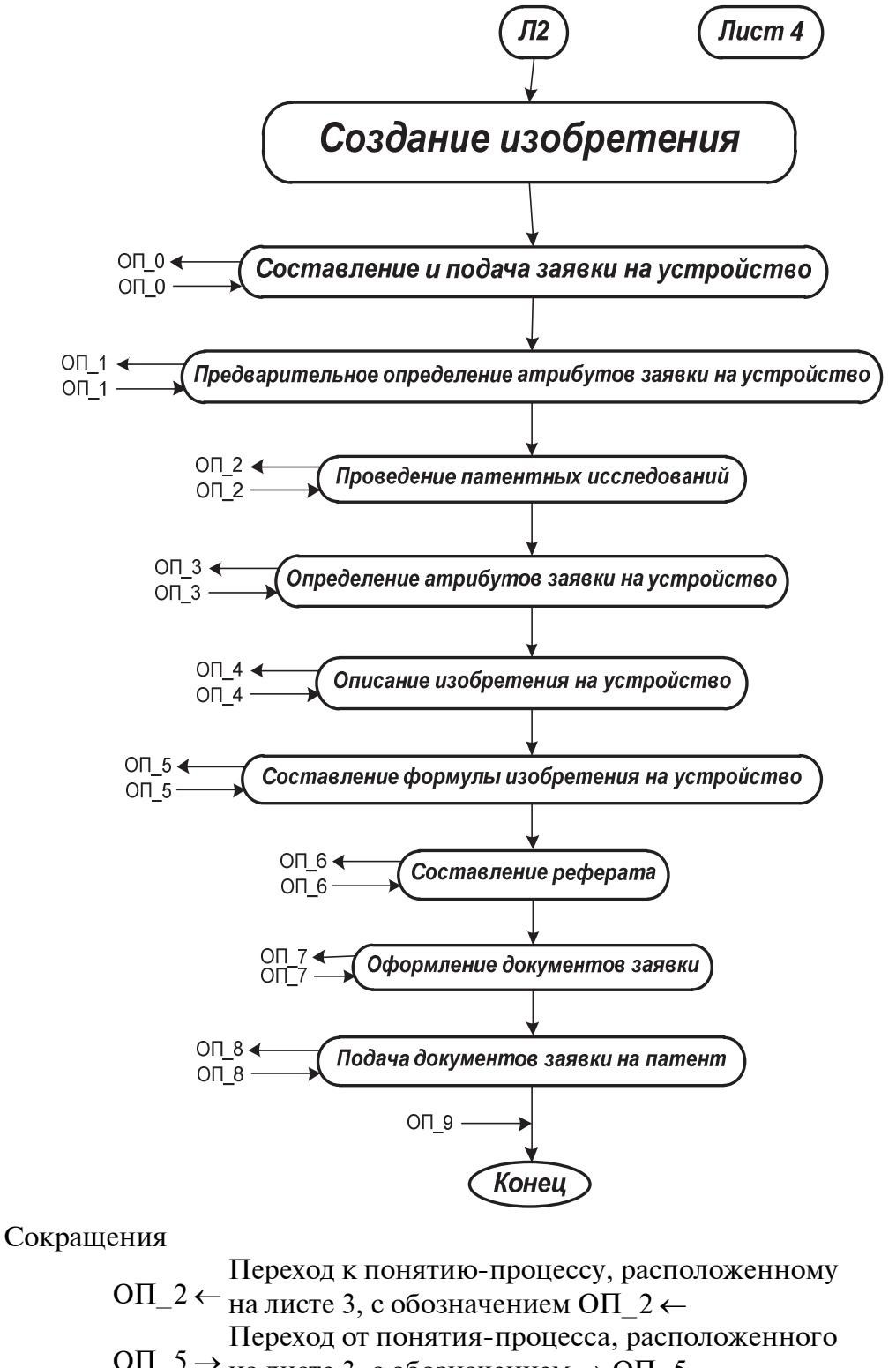

***Рис. 8.*** Онтология задач создания и оформления изобретения на устройство

## Заключение

Методологические основы построения систем ИП, разработанные на основе концепций трансдисциплинарности и онтологического управления, включают модели, технологию и инструментальные средства когнитивных процессов обработки информации, в том числе проектирования объектов новой техники, автоматизированного построения онтологических баз знаний ПдО. Последние есть базовыми компонентами интеллектуальных технологий и систем при проведении сложных научных исследований междисциплинарного и трансдисциплинарного характера. Технология системной интеграции онтологий ПдО позволяет исследовать взаимодействие предметных знаний как на стыке предметных дисциплин, так и в кластерах конвергенции, что открывает широкие возможности для получения новых знаний и для разработки научных теорий. В статье рассмотрен пример создания и патентования лингвистического процессора для обработки больших объемов текстовой информации с целью последующего извлечения знаний.

СПИСОК ЛИТЕРАТУРЫ

*Aleksandr Palagin*, Academician of NAS of Ukraine (Computer Science), Professor,
Doctor of Technical Sciences (Computers, Systems and Networks), deputy director
palagin_a@ukr.net

*Nykolay Petrenko*, Doctor of Technical Sciences, Leading Researcher, Microprocessor Technology Department,
petrng@ukr.net

*Kyryl Malakhov*, Master of Information Technology, Junior Researcher, Microprocessor Technology Department,
malakhovks@nas.gov.ua

V.M. Glushkov Institute of Cybernetics, The National Academy of Sciences of Ukraine,
Glushkov ave., 40, Kyiv, 03187, Ukraine

## INFORMATION TECHNOLOGY AND INTEGRATED TOOLS FOR SUPPORT OF SMART SYSTEMS RESEARCH DESIGN

**Introduction.** The article describes the methodological foundations of the development of scientific research design systems, which include: models, technology and tools for processing large volumes of text information/data; the formation of conceptual structures (ontologies) of different branches of knowledge, their system integration based on transdisciplinary and ontology concepts.

**Purpose.** The purpose of the article is to develop information technology for the smart systems design.

**Methods.** The methods and models used for the smart systems design are based on the transdisciplinary and ontological concepts of knowledge information processing. Ontologies are the core components of intelligent technologies and systems in comprehensive interdisciplinary and transdisciplinary scientific research. The process of system integration of ontologies makes it possible to investigate the interaction of object knowledge, as an object at the intersection of disciplines, as well as in clusters of convergence, which opens opportunities for new knowledge and for the development of scientific theories. Integrated tools implement a unified information technology, which is based on the proposed function, information and behavioral UML-models. Information technology is designed to support the scientific research design of smart systems, which is a type of scientific research and is characterized by the fact that its main stages are related to the process of describing the alleged pattern of the projected object of the new engineering. The design process itself is built as a series of interactive procedures of deep immersion into the description of the new engineering object and the formation of the intermediate variants of the technical solution. In this case, the transformation of the object design description depends, both on object knowledge in general, and on the knowledge and experience of the designer.

**Results.** The information technology of research-related design of smart systems is developed. An example of the creation and patenting of a linguistic processor for processing large amounts of text data with the purpose of subsequent extraction of knowledge is considered.

**Conclusion.** The methodological foundations of the scientific research design systems development include models, technology, and tools for cognitive processes of information processing, including the design of the new engineering objects, the automated development of ontological knowledge bases of the knowledge domain. The last ones are the basic components of intelligent technologies and systems for comprehensive scientific research of the interdisciplinary and transdisciplinary nature. The process of system integration of ontologies makes it possible to investigate the interaction of object knowledge, as an object at the intersection of disciplines, as well as in clusters of convergence, which opens opportunities for new knowledge and the development of scientific theories.

*О.В. Палагін*, академік НАНУ, д-р техн. наук, професор, заступник директора, palagin_a@ukr.net,

*М.Г. Петренко*, д-р техн. наук, ведуч. наук. співроб., petrng@ukr.net,

*К.С. Малахів*, молодш. наук. співроб., malakhovks@nas.gov.ua,

Інститут кібернетики імені В.М. Глушкова НАН України, просп. Глушкова, 40, Київ, 03187, Україна

## ІНФОРМАЦІЙНА ТЕХНОЛОГІЯ ТА ІНСТРУМЕНТАЛЬНІ ЗАСОБИ ПІДТРИМКИ ПРОЦЕСІВ ДОСЛІДНОГО ПРОЕКТУВАННЯ *SMART*-СИСТЕМ

**Вступ.** Розглянуто методологічні основи побудови систем дослідного проектування, які містять моделі, технологію та інструментальні засоби процесів обробки великих обсягів текстової інформації, формування концептуальних структур (онтологій) різних галузей знань та їх системної інтеграції.

**Мета статті —** розробка інформаційної технології проектування *Smart*-систем.

**Методи:** методи і моделі, використані в розробці, ґрунтуються на трансдисциплінарній та онтологічній концепціях роботи зі знаннями. Онтології є базовими компонентами інтелектуальних технологій та систем за проведення складних наукових досліджень міждисциплінарного і трансдисциплінарного характеру. Процес системної інтеграції онтологій дозволяє досліджувати взаємодію предметних знань як на межі предметних дисциплін, так і в кластерах конвергенції, що відкриває широкі можливості для отримання нових знань та розробки наукових теорій. Інструментальні засоби реалізують єдину інформаційну технологію, в основу якої покладено запропоновані функціональна, інформаційна та поведінкова моделі, що базуються на *UML*-діаграмах. Інформаційна технологія призначена, в тому числі, для підтримки процесів дослідного проектування *Smart*-систем, яке є різновидом наукових досліджень і характеризується тим, що основні його етапи пов'язані з процесом опису можливого образу проектованого об'єкта нової техніки. Власне процес проектування будується як низка інтерактивних процедур більш глибокого занурення в опис об'єкта нової техніки і формування проміжних варіантів технічного рішення. При цьому перетворення опису об'єкта проектування залежить як від знань про предметну область в цілому, так і від знань та досвіду проектувальника.

**Результат.** Розроблено інформаційну технологію дослідного проектування *Smart*-систем. Наведено приклад створення і патентування лінгвістичного процесора для обробки великих обсягів текстової інформації з метою подальшого отримання знань.

**Висновок.** Методологічні основи побудови систем дослідного проектування містять моделі, технологію та інструментальні засоби когнітивних процесів обробки інформації, а також проектування об'єктів нової техніки, автоматизованої побудови онтологічних баз знань предметної області. Останні є базовими компонентами інтелектуальних технологій і систем при проведенні складних наукових досліджень міждисциплінарного і трансдисциплінарного характеру. Технологія системної інтеграції онтологій ПдО дозволяє дослідити взаємодію предметних знань як на межі предметних дисциплін, так і в кластерах конвергенції, що відкриває широкі можливості для отримання нових знань та розробки наукових теорій.